# High-efficiency perovskite-polymer bulk heterostructure light-emitting diodes

Baodan Zhao[1], Sai Bai[2,3], Vincent Kim[1], Robin Lamboll[1], Ravichandran Shivanna[1], Florian Auras[1], Johannes M. Richter[1], Le Yang[1,4], Linjie Dai[1], Mejd Alsari[1], Xiao-Jian She[1], Lusheng Liang[5], Jiangbin Zhang[1], Samuele Lilliu[6,7], Peng Gao[5], Henry J. Snaith[2], Jianpu Wang[8], Neil C. Greenham[1], Richard H. Friend[1]* & Dawei Di[1]*

1. Cavendish Laboratory, Cambridge University, JJ Thomson Avenue, Cambridge, CB3 0HE, UK
2. Department of Physics, University of Oxford, Clarendon Laboratory, Oxford, OX1 3PU, UK
3. Department of Physics, Chemistry and Biology (IFM), Linköping University, Linköping, SE-581 83, Sweden
4. Institute of Materials Research and Engineering (IMRE), Agency for Science, Technology and Research (A*STAR), 2 Fusionopolis Way, Singapore 138634, Singapore
5. Laboratory of Advanced Functional Materials, Xiamen Institute of Rare Earth Materials, Haixi Institute, Chinese Academy of Sciences, Xiamen 361021, China
6. Department of Physics and Astronomy, University of Sheffield, Sheffield, S3 7RH, UK
7. The UAE Centre for Crystallography, United Arab Emirates
8. Institute of Advanced Materials (IAM), Nanjing Tech University, 30 South Puzhu Road, Nanjing, 211816, China

* Email: dd403@cam.ac.uk (D.D.); rhf10@cam.ac.uk (R.H.F.).

**Abstract:**

**Perovskite-based optoelectronic devices have gained significant attention due to their remarkable performance and low processing cost, particularly for solar cells. However, for perovskite light-emitting diodes (LEDs), non-radiative charge carrier recombination has limited electroluminescence (EL) efficiency. Here we demonstrate perovskite-polymer bulk heterostructure LEDs exhibiting record-high external quantum efficiencies (EQEs) exceeding 20%, and an EL half-life of 46 hours under continuous operation. This performance is achieved with an emissive layer comprising quasi-2D and 3D perovskites and an insulating polymer. Transient optical spectroscopy reveals that photogenerated excitations at the quasi-2D perovskite component migrate to lower-energy sites within 1 ps. The dominant component of the photoluminescence (PL) is primarily bimolecular and is characteristic of the 3D regions. From PL quantum efficiency and transient kinetics of the emissive layer with/without charge-transport contacts, we find non-radiative recombination pathways to be effectively eliminated. Light outcoupling from planar LEDs, as used in OLED displays, generally limits EQE to 20-30%, and we model our reported EL efficiency of over 20% in the forward direction to indicate the internal quantum efficiency (IQE) to be close to 100%. Together with the low drive voltages needed to achieve useful photon fluxes (2-3 V for 0.1-1 mA cm$^{-2}$), these results establish that perovskite-based LEDs have significant potential for light-emission applications.**

**Main Text:**

The rapid advance of perovskite solar cells[1–4] has prompted the development of other types of perovskite-based optoelectronics, including LEDs[5–8], lasers[9,10] and photo-detectors[11,12]. Similar to conventional semiconductors, the luminescence efficiency of photo- or electrically-excited charge carriers in the perovskite material is governed by the relative strengths of radiative and non-radiative processes. Since the first report of halide perovskite LEDs in 2014[5], the device EQEs have risen from ~0.1%[5] to ~14%[13]. While these high EL efficiencies are surprising for the solution-processed hybrid semiconductor containing a considerable level of grain-boundary and interfacial defects, such device performance still falls behind that of the best OLEDs[14–16] which exhibit EQEs of more than 20% without enhanced optical outcoupling. Despite the good tolerance of the hybrid perovskite materials family to electronic defects[17], the EL efficiencies achieved to date suggest that the suppression of non-radiative recombination under electrical excitation conditions still remains a challenge. Non-radiative recombination is also an important mechanism for voltage loss in photovoltaic solar cells[18]. An ideal open-circuit voltage predicted by the Shockley-Queisser model[19] is only achievable with near-unity luminescence yield[20]. To enhance radiative emission processes in LEDs, one of the most successful approaches is the use of low-dimensional structures such as nanocrystals[21–23] and quasi-2D/3D nanostructures[7,8] that are considered to confine charge carriers.

In this paper, we explore the optoelectronic and photophysical properties of a perovskite-polymer bulk heterostructure (PPBH). The emissive heterostructure is prepared from a combination of quasi-2D/3D perovskites and a wide optical gap polymer (Fig. S1). The absorption profile of the PPBH contains a distinct excitonic peak at ~575 nm, corresponding to the quasi-2D perovskite with a formula of $A_2B_{m-1}Pb_mI_{3m+1}$, where A and B are organic cations (Fig. 1a). The absorption tail of the PPBH sample extends to ~800 nm, which we attribute to the quasi-3D phase. The photoluminescence (PL) spectrum of the sample peaks at ~795 nm (~1.56 eV), with a full-width-at-half-maximum (FWHM) of ~55 nm. Grazing-incidence wide-angle X-ray scattering (GIWAXS) measurements indicate that the perovskite crystallites are isotropically oriented in the PPBH film (Fig. 1b). High-resolution transmission electron microscopy (HR-TEM) results suggest the presence of quasi-2D/3D crystal structures[24] (Fig. 1c). From the X-ray diffraction (XRD) data, the average crystallite size is estimated to be 30-55 nm based on the FWHM of the diffraction peaks (Fig. S2a). The average surface roughness of the film is ~3.3 nm (Fig. S2c).

To investigate the electroluminescence (EL) properties of the PPBH, we developed a solution-processed multilayer LED structure (Fig. 1d). The EL spectrum of the PPBH LED is nearly identical to that of the steady-state PL (Fig. S2d), exhibiting a slightly narrower FWHM of ~49 nm. The low onset of radiance ($10^{-4}$ W sr$^{-1}$ m$^{-2}$ at 1.3 V) indicates that the LED structure allows barrier-free bi-polar charge injection into the emissive layer (Fig. 1e). The peak EQE of best devices reaches 20.1% (Fig. 1f), representing a record for perovskite-based LEDs (Fig. S3)[5–8,22,25]. The angle-dependent EL intensities exhibit a Lambertian profile (Fig. S4a), allowing accurate estimation of EQEs. As the drive voltages are low, the wall-plug efficiency (electricity-to-light power-conversion efficiency) is also high, reaching 16.2% (Fig. S4b). The efficiencies of our perovskite-based LEDs are on par with those of the best OLEDs[14–16] and quantum-dot (QD) LEDs[26]. For devices encapsulated in epoxy adhesive/cover glass, the EL half-life in air under continuous operation at the current density corresponding to the peak EQE point (0.1 mA cm$^{-2}$) has reached 46 hours (Fig. 1g). Although practical applications demand further improvements, the half-life of our devices represents the longest operational lifetime observed for perovskite-based LEDs to date[8,23].

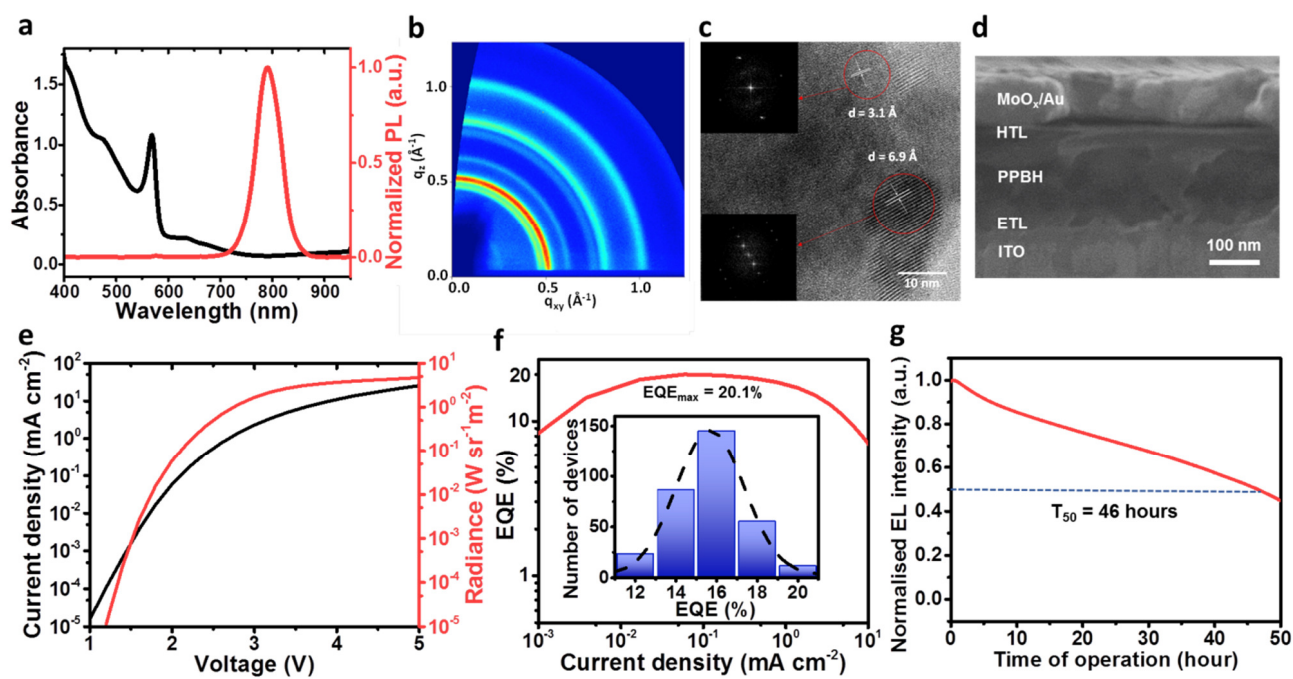

**Figure 1 | Basic characterisation of PPBH and LED performance. a**, Absorbance (black) and PL (red) spectra of a PPBH film on fused silica. **b**, GIWAXS patterns of a PPBH layer deposited on silicon. **c**, HR-TEM image of a PPBH sample. Insets show the fast Fourier transforms of the quasi-2D/3D crystalline regions. **d**, Cross-sectional scanning electron microscopy (SEM) image of the LED structure. **e**, Current-voltage and radiance-voltage characteristics. **f**, EQE-current density characteristics of the best PPBH LED (peak EQE=20.1%) and peak EQE histogram of 320 devices (inset). **g**, Operational stability measurement of PPBH LEDs, performed in air at a constant current density of 0.1 mA cm$^{-2}$ corresponding to the peak EQE point.

For films of PPBH (thickness: ~180 nm) deposited on fused silica substrates, we observe external PL quantum efficiencies (PLQEs) of 90±10% under 532-nm steady-state excitation (10-300 mW cm$^{-2}$) using an integrating sphere. To translate the high PLQE into good EL efficiency, a key consideration is the prevention of luminescence quenching at the charge-transport interfaces, as interfacial states often create non-radiative pathways, imposing severe limitations on optoelectronic performance[27]. Accurate assessment of PLQE of the emissive layer in an LED structure using an integrating sphere is limited by the parasitic absorption of charge-transport layers and electrodes. Therefore, we employ a range of transient optical spectroscopy techniques in this work to investigate excitation and recombination processes in the PPBH system. We find that the charge-transport layers have no observable effect on the PL decay kinetics of the PPBH emissive layer (Fig. 2a and Fig. S5a), indicating that non-radiative recombination events at the two contact interfaces are insignificant. In contrast, for identically-prepared quasi-2D/3D perovskite films without the polymer component, bulk and interfacial luminescence quenching processes are clearly present (Fig. S5b-d). Fig. 2b shows the transient PL and EL kinetics of a PPBH LED. We do not observe any evidence for non-radiative decay processes in the transient EL profile, similar to what has been found in the PL kinetics, consistent with balanced charge injection and efficient operation.

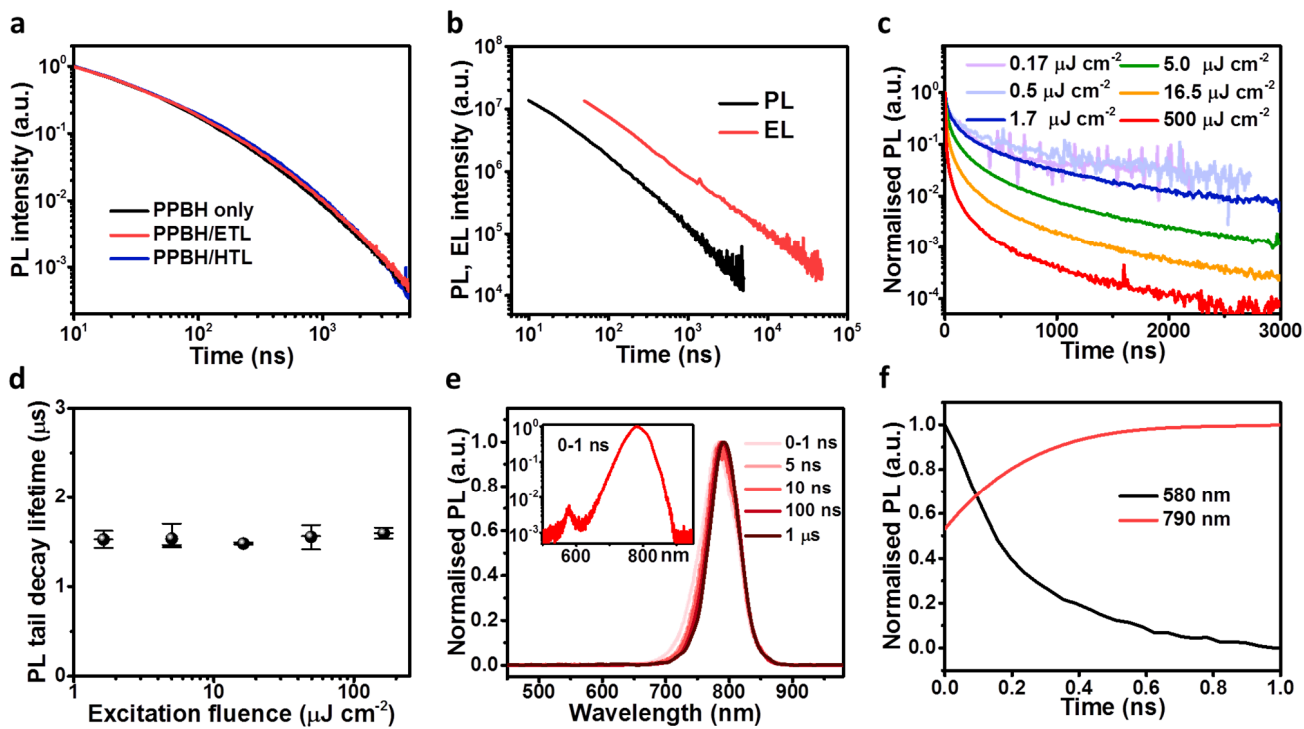

**Figure 2 | Transient (ns-µs) optical experiments. a**, PL kinetics of a PPBH film and identical PPBH samples with charge-transport layers. **b**, Transient PL and EL measurements of a PPBH LED. The PL and EL decays show comparable initial intensities. The PL decay was measured at 10-ns time-steps. During the transient EL measurements, the device was driven by 2 V/0 V (on/off) square voltage pulses at a frequency of 1 kHz. The EL decay was recorded during the off-cycles of the pulses, with a 50-ns time-step. **c**, Excitation-intensity-dependent PL kinetics of a PPBH film. The initial intensities of all PL decay trances are normalised to 1 at $t = 0$. **d**, PL decay tail lifetime as a function of excitation intensity. **e**, ns-µs time-resolved PL spectra. The excitation source for the PL experiments in Fig. 2a-e was a 400-nm, 80-fs pulsed laser with a repetition frequency of 1 kHz. Unless otherwise specified, the excitation fluence of the PL measurements was 5 µJ cm$^{-2}$. An intensified CCD (iCCD) system with a temporal resolution of ~2 ns was used. **f**, Instrumentally-limited kinetics of the short (~580 nm) and long (~790 nm) wavelength PL components measured by time-correlated single photon counting (TCSPC). The excitation source was a 407-nm laser with a pulse width of < 200 ps. The temporal resolution of the setup is ~0.2 ns.

The excitation-fluence-dependent PL decay profiles of the PPBH sample from 0 ns to 3 µs at 10-ns time-steps are shown in Fig. 2c. The PL decay kinetics show a significant dependence on excitation intensity, consistent with the view that in the emissive layer, the carrier recombination rate can be understood by a superposition of first-order (monomolecular), second-order (bimolecular) and higher-order processes, as has been discussed in earlier reports[9,28,29]. The early-time emission process in the PPBH system is dominated by an energy-migration process (*vide infra*), and hence deviates from this model.

We find that for the PL process at 5 ns < $t$ < 1 µs, which accounts for ~90% of the total emission for excitation fluences greater than 1.7 µJ cm$^{-2}$, the decay kinetics can be approximately described by a bimolecular process. At later times ($t$ > 1 µs), when the carrier density in the system is low, the PL decays follow a mono-exponential process. The lifetimes of these PL decay tails are approximately 1.5 µs for a range of excitation intensities (Fig. 2d). This lifetime is an indication of trap-assisted non-radiative recombination lifetime of photoexcitations at low carrier densities, or the intrinsic decay lifetime of weakly bound electron-hole pairs in the PPBH.

From the ns-μs PL spectra of the PPBH (Fig. 2e), we observed that the PL profile stabilised to the steady-state PL after $t > 10$ ns, following an initial spectral redshift. At early times ($t < 1$ ns), a weak PL contribution from the quasi-2D perovskite was observed. Accurate determination of the decay and rise times of the high- (~580 nm) and low-energy (~790 nm) emissive species was limited by the temporal response (0.2 ns) of the experimental setup (Fig. 2f). Ultrafast PL experiments in the fs-ps time-range (Fig. 3a) show that photoexcitation at 400 nm to mostly the quasi-2D perovskite gives early-time emission from these regions (peaked at ~580 nm), with a characteristic lifetime of ~10 ps (Fig. 3b), and is insensitive to excitation intensity (Fig. S6a,b). The fs-ps PL kinetics of the longer-wavelength (> 680 nm) component could not be recorded by our ultrafast PL setup due to the limited spectral detection window[16].

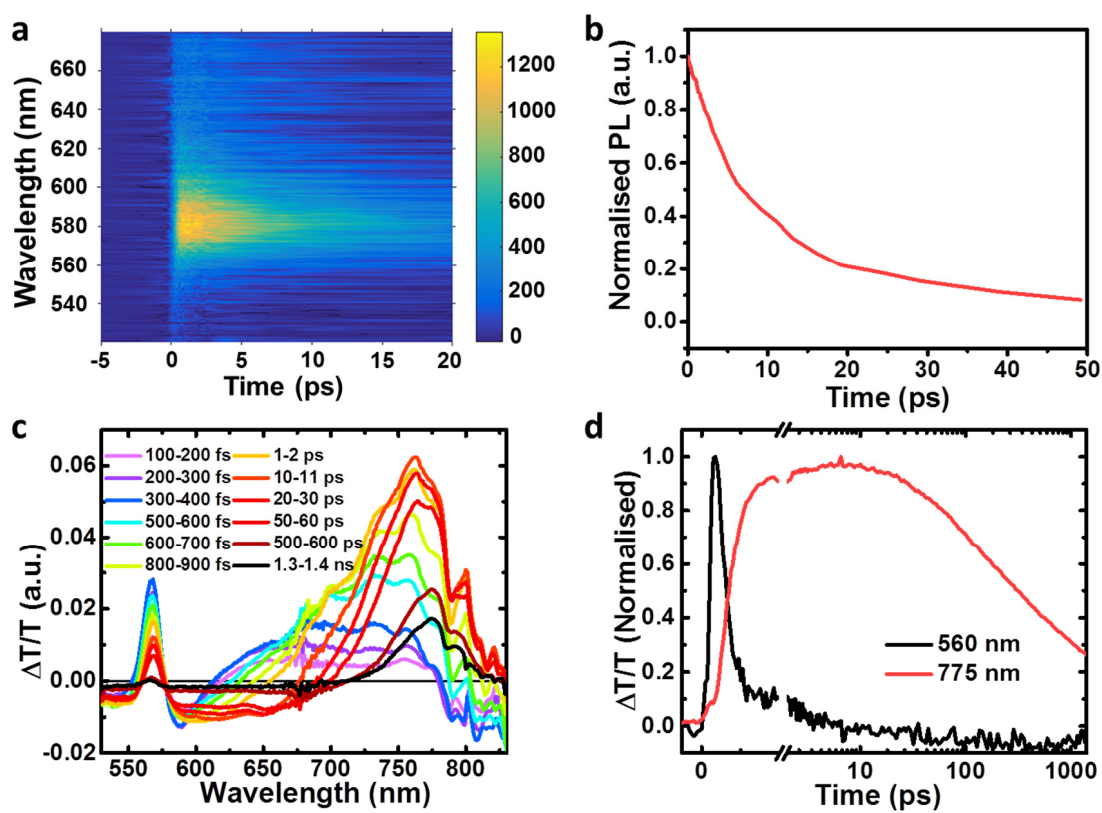

**Figure 3 | Ultrafast (fs-ps) transient optical experiments. a**, Ultrafast (ps) PL spectra. The excitation fluence used was 5 μJ cm⁻². **b**, Ultrafast (ps) PL decay kinetics at 550-620 nm. **c**, Transient absorption (TA) spectra of the sample. The excitation fluence for the TA measurement was 16 μJ cm⁻². **d**, TA kinetics of the high- and low-energy absorption species. For all ultrafast optical measurements, the excitation source was a 400-nm, 80-fs pulsed laser with a repetition rate of 1 kHz.

Transient absorption (TA) experiments (Fig. 3c) confirm that the initial photoexcitation is formed on the quasi-2D perovskite, with a ground-state bleach (GSB) peak at ~575 nm. The peak position is in agreement with the steady-state absorption and transient PL measurements. The GSB feature related to the quasi-2D perovskite has an exciton-like character, suggesting the electron-hole pair created in the material is initially bound. As the 575-nm GSB quickly decays, a red-shifting, broad GSB signal rises rapidly and then declines after 10 ps. The kinetics of the ~575-nm and the ~775-nm processes which are related to the emissive species identified by the PL experiments are shown in Fig. 3d. The data show that the initial excitation at the 575-nm high-energy site relaxes within ~1 ps, while the remaining states with the same energy decay in ~10 ps, consistent with the ultrafast PL experiment showing a 10-ps decay time. The 1-ps rise time of the lower-energy (~775 nm) excitations suggests that the energy transfer from higher-bandgap to lower-bandgap components in the PPBH takes place within the same timeframe. Following the initial 1-ps ultrafast energy migration, the residual localised excitations at the quasi-2D perovskite decay in ~10 ps. The majority of the excited states in the lower-bandgap components likely exist in the form of charged excitations rather than tightly-bound excitons, as implicated by the broadened GSB features of the lower-energy excitations. At later times ($t > 1$ ns), most of these charge carriers recombine to produce luminescence, as discussed previously in the ns-μs transient studies.

Finally, we discuss models of photon extraction from the LED structure. From a conventional optics perspective, we note that the refractive index of the PPBH system which we measure to be 1.9 at the peak of its steady-state luminescence is significantly lower than the refractive index of 2.7 for standard halide perovskites. This lower refractive index widens the escape cone of photon emission from the emissive layer to 32°. Including the effects of interference and assuming experimentally-determined optical constants and layer thicknesses, we model an outcoupling factor of up to ~21% (Fig. 4a), depending on where in the emissive layer photons are emitted, with values of ~25% possible if emissive layer thicknesses are optimised (Fig. 4b). The similarity between modelled outcoupling factor and EQE implies near-unity IQEs, in agreement with the outstanding PL and EL properties observed. Furthermore, lateral PL experiments (Fig. 4c, d and Fig. S7) show that light initially confined as modes waveguided in PPBH layer can propagate up to 80 μm. The intense PL signal at $d < 10$ μm resembles the intensity profile of the excitation laser spot. The PL decay beyond 10 μm (the system resolution) has a characteristic decay profile similar to that observed previously in halide perovskites[30], indicating a possible contribution from photon-recycling. Therefore, while an EQE of over 20% for PPBH LEDs can be accounted for within the conventional outcoupling model described above, photon recycling provides prospects of further efficiency improvements.

In this work, we have demonstrated PPBH LEDs with high EQEs of up to 20.1% and an EL half-life of 46 hours. They represent the most efficient perovskite-based LEDs to date, and are comparable with some of the best OLEDs and QD LEDs[14–16,26]. Transient optical experiments suggest that in the PPBH system, localised higher-energy excitations dissociate into charges at lower-energy sites within ~1 ps, significantly faster than the ~100 ps energy funneling time reported for quasi-2D/3D perovskites[7,8]. The ultrafast migration of excitations observed ensures that any non-radiative traps with energies above the final emissive species are rendered insignificant, as these loss processes need to compete kinetically with the rapid excitation transfer. After the rapid energy migration, emission occurs primarily through a bimolecular recombination channel. PPBH thin films with/without charge-transport contacts exhibit a PL decay tail (monomolecular) lifetime of ~1.5 μs, comparable to or exceeding that observed for perovskite single crystals[31,32], indicating low trap densities in these structures. Bulk and interfacial non-radiative relaxation processes are effectively suppressed, leading to the excellent EQEs and near-100% PLQEs. Optical outcoupling from the emissive layer is improved by the reduced effective refractive index, and our measured EQE is consistent with this conventional model for thin-film LEDs. However, low absorption losses for modes guided in-plane may enable performance enhancement through photon recycling that allows outcoupling from these modes[30]. We anticipate the outstanding optoelectronic properties of the PPBH system to lead to low-cost and high-performance photon sources for lighting, display and communication technologies.

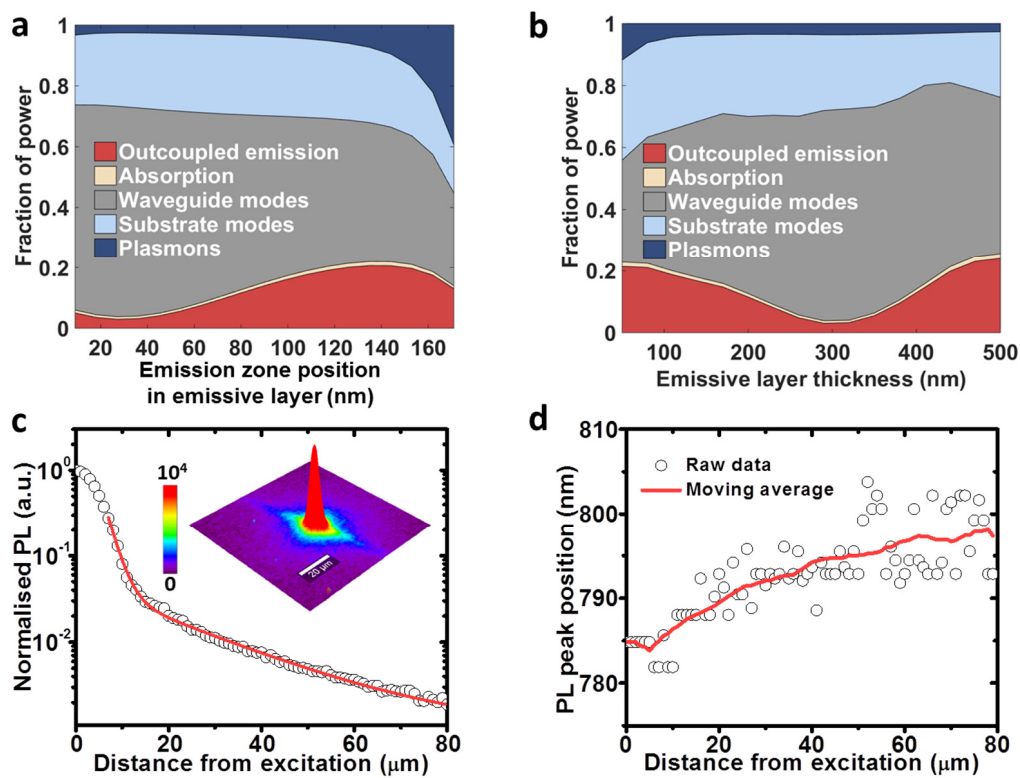

**Figure 4 | Optical modeling of PPBH LEDs and lateral PL experiments. a**, Modelled fractional optical power distribution in the LED structure as a function of emission zone position in the PPBH emissive layer. The origin of the x-axis is the ETL/emissive layer interface. 'Outcoupled emission' indicates the fraction of out-coupled light from the LED. Other modes lead to optical losses. **b**, Modelled fractional optical power as a function of PPBH layer thickness. The emission zone is assumed to be at the centre of the emissive layer. **c**, PL intensity as a function of lateral distance from the point of excitation. **d**, PL peak position as a function of lateral distance from the point of excitation. The redshift of PL peak over distance indicates spectrum filtering (self-absorption) within the emissive layer.

## Acknowledgements

B.Z. thanks the Cambridge Trust and China Scholarship Council for funding and support. S.B. is supported by a VINNMER Marie-Curie Fellowship. R.S. acknowledges the Royal Society Newton-Bhabha International Fellowship. M.A. acknowledges the President of the UAE's Distinguished Student Scholarship Program (DSS), granted by the UAE's Ministry of Presidential Affairs. XMaS is a mid-range facility supported by the Engineering and Physical Sciences Research Council (EPSRC). We are grateful to all the XMaS beamline team staff for their support. P.G. acknowledges the 'Thousand Talent Program' for support. D.D. and R.H.F. acknowledge the EPSRC for support.

## Author contributions

B.Z. and D.D. developed and characterised the high-efficiency LEDs. D.D., B.Z. and L.Y. carried out the ns-μs transient PL and EL studies. V.K. conducted the transient absorption experiments. J.M.R. performed the fs-ps transient PL measurements. R.L. developed the optical model for the LED devices under the guidance of N.C.G. S.B. synthesized the nanocrystals, tailored the nanocrystal properties and contributed to the LED development. R.S. carried out the lateral PL experiments. F.A., L.L. and P.G. synthesized the perovskite precursors. L.D. performed the HR-TEM and AFM measurements. X.-J.S. and B.Z. performed the SEM studies. F.A., M.A., P.G. and B.Z. carried out the XRD analysis. J.Z. helped with some transient measurements and PLQE calculations. D.D. and B.Z. analysed all the results and wrote the manuscript which was revised by R.H.F. All authors contributed to the work and commented on the paper. D.D. planned the project and guided the work with R.H.F.

## Competing financial interests

The authors declare no competing financial interests.

## Corresponding authors

Dawei Di (dd403@cam.ac.uk); Richard H. Friend (rhf10@cam.ac.uk).